\documentclass[aps,prd,preprint,preprintnumbers,unsortedaddress,superscriptaddress,showpacs,nofootinbib]{revtex4-1}
\pdfoutput=1
\usepackage{graphicx}
\usepackage{amsmath}
\usepackage{amsfonts}
\usepackage{amssymb}
\usepackage{color}

\usepackage[colorlinks, citecolor=blue,anchorcolor=red,menucolor=red, linkcolor=red,filecolor=red,runcolor=red,urlcolor=blue,frenchlinks=red]{hyperref}

\usepackage{bm}
\usepackage[section]{placeins}
\usepackage{booktabs}%改表格线条粗细hline, 使用\toprule, \midrule and \bottomrule
\newcommand{\blue}[1]{\textcolor{blue}{#1}}

\begin{document}%\bibliographystyle{unsrt}
\arraycolsep1.5pt

\title{Helicity amplitudes in the $\bar{B} \to  D^{*} \bar{\nu}_\tau \tau$  decay\\ with $V-A$  breaking  in the quark sector}

%\author{L.~R.~Dai~\footnote{Corresponding author}}

\author{L.~R.~Dai}
\email{dailr@lnnu.edu.cn}
\affiliation{Department of Physics, Liaoning Normal University, Dalian 116029, China}

\author{E. Oset}
\email{oset@ific.uv.es}
\affiliation{Departamento de F\'isica Te\'orica and IFIC, Centro Mixto Universidad de Valencia-CSIC,
Institutos de Investigac\'ion de Paterna, Aptdo. 22085, 46071 Valencia, Spain
}

\date{\today}
\begin{abstract}
In view of the recent measurement of the $F_L^{D^*}$ magnitude in the   $\bar{B} \to  D^{*} \bar{\nu}_\tau \tau$  reaction we evaluate this magnitude within the standard model
and for a family of models with the $\gamma^\mu -\alpha\gamma^\mu \gamma_5$ current structure for the quarks for different values of $\alpha$. At the same time we evaluate also
the transverse contributions, $M=-1$, $M=+1$, and find that the difference  between the  $M=-1$ and  $M=+1$ contributions is far more sensitive to changes in $\alpha$ than
the longitudinal component.
 \end{abstract}

\maketitle

\section{Introduction}

The measurement of vector polarization in $B$ decays with vectors in the decay products has captured  the attention of the physics community as promising tools of information on
possible physics beyond the standard model (BSM) \cite{aubert,data,data2,altonen,cldamo,lhcb,kagan,beneke,data3,xuwang,lees,belle1,cdf,cms,aaij,gudrum,buras,lu1,lu2,lu3}.  In particular the
information  on the helicity amplitudes in the $B \to  D^{*} \bar{\nu} l$ and $\bar{B} \to  D^{*} \bar{\nu}_\tau \tau$  has been advocated in \cite{fajfer,tanaka} as useful tools to explore
physics  BSM.

In a recent paper \cite{belle} the Belle Collaboration has reported the first measurement  of the $D^*$ polarization in the $\bar{B} \to  D^{*} \bar{\nu}_\tau \tau$  decay,
providing a value for $F_L^{D^*}$, the fraction of the  longitudinal  polarization contribution to the total,  of
 \renewcommand{\theequation}{1}
 \begin{eqnarray}
F_L^{D^*}=0.60\pm 0.08 \pm 0.04 \,.
\end{eqnarray}

In the standard model (SM), $F_L^{D^*}$ is about $0.45$ \cite{bstd,ivanov,alok,huang,srimoy}, with the most recent predictions giving $0.441 \pm 0.006 $ \cite{huang}
and $0.457 \pm 0.010 $ \cite{srimoy} and $0.476^{+0.015}_{-0.014}$ \cite{new}.  Models BSM with scalar and tensor contributions can produce sizeable changes in $F_L^{D^*}$, as shown in  \cite{bstd,alok,huang,srimoy,ivanov2}.

In the present work we want to show that for some family of models BSM the fraction $F_L^{D^*}$ is rather insensitive to changes from the SM, while  the transverse
polarizations are more sensitive and  in particular  the difference between the $M'=-1$  and $M'=+1$ components is very sensitive to these changes.

In a former paper  \cite{epjc78} we studied the helicity amplitudes of the $B \to  D^{*} \bar{\nu} l$  transition for a model in which the quark current is given by
 \renewcommand{\theequation}{2}
 \begin{eqnarray}
Q^\mu=\langle {\bar u}_c|\gamma^\mu(1-\blue{\alpha}\gamma_5)|u_b\rangle \,,
\end{eqnarray}
 where $\alpha=1$ for the SM.  The calculations are done using a quark model for the operators \cite{first}, with a mapping of the quark momenta to those of the mesons consistent with
 heavy quark symmetry \cite{isgur,wise,neubert}.  The longitudinal, $M'=0$, and two transverse  $M'=-1$,  $M'=+1$ polarization  contributions, taking the $z$ axis along the $D^*$ direction in the
$\bar{\nu} l$  rest frame, were evaluated. It was found that for  different  values of $\alpha$ the magnitude most sensitive to the change was the difference
between the $M'=-1$  and $M'=+1$ contributions.

   The model of  \cite{epjc78} neglects the contribution of intrinsic quark form factors, which are claimed to approximately cancel in the ratios evaluated  there.

In view of the measurement of \cite{belle}, we find most opportune to test this model with this measurement, extending our model  of the   $B \to  D^{*} \bar{\nu} l$ reaction to the
$\bar{B} \to  D^{*} \bar{\nu}_\tau \tau$  one.

\section{Formalism}
\begin{figure}[ht]
\includegraphics[scale=0.72]{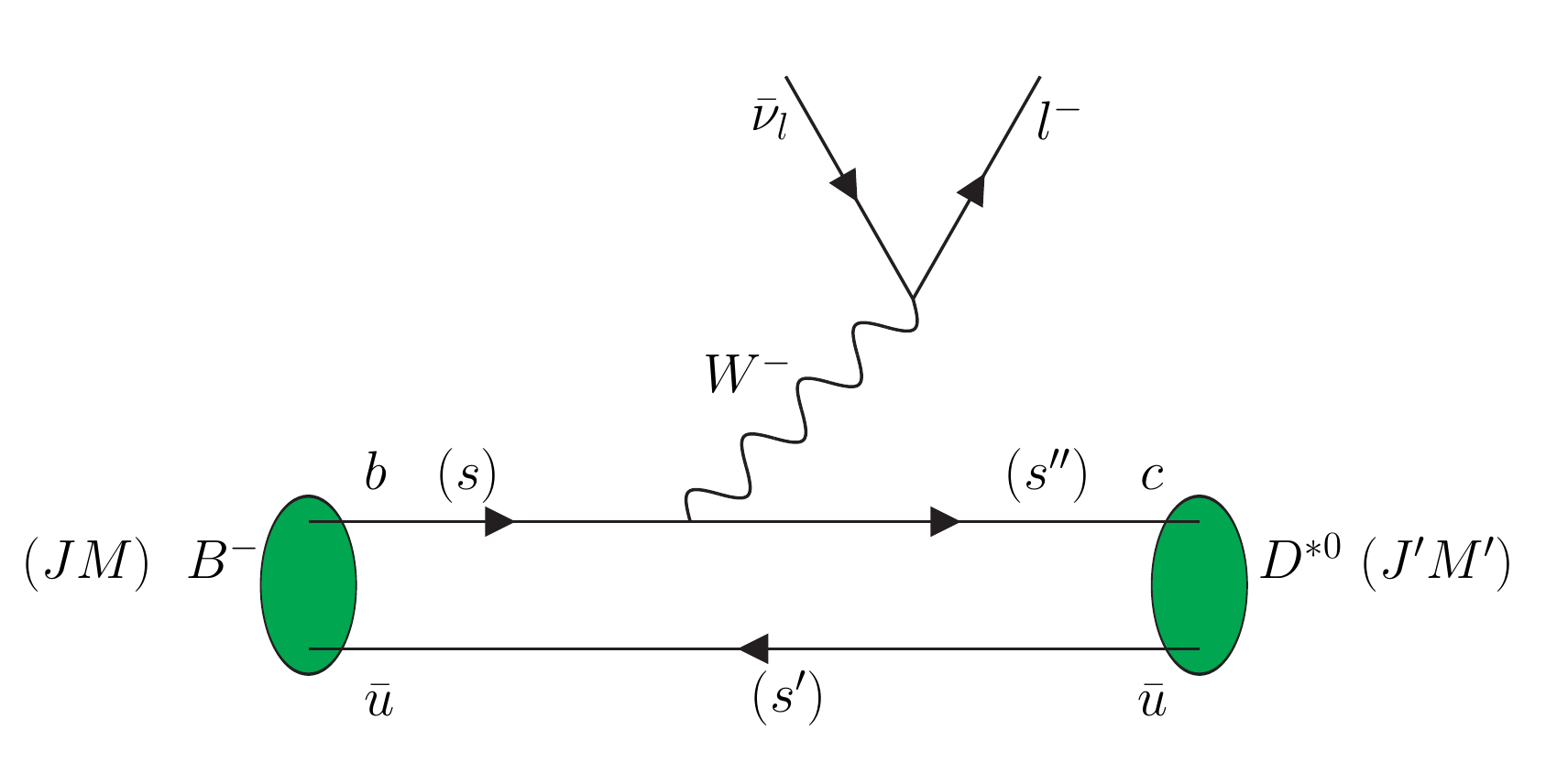}
\caption{Diagram of  $B^- \to  D^{*0} \bar{\nu}_{l} l^-$  at the quark level.}
\label{fig:diag}
\end{figure}

In the present work we will  study the $B \to  D^{*} \bar{\nu} l$  decay, which  is  depicted  in Fig. \ref{fig:diag}  for  $B^- \to D^{*0} \bar{\nu}_{l} l^-  $.

We use the same  nomenclature as in \cite{epjc78} where a study of the meson decays $J M  \to  \bar{\nu}_{l} l J' M'$  was done, where $ J M (J' M')$
 are the modulus and third component of the initial (final) meson spin, and the rates for the different third components in the $J=0, J'=1$ case were evaluated.

The differential width  for  $B \to  D^{*} \bar{\nu} l$    is given   by
  \renewcommand{\theequation}{3}
\begin{eqnarray}\label{eq:fac18}
\frac{d \Gamma}{d M_{\rm inv}^{(\nu l)}}= \frac{2m_{\nu} 2m_l}{(2\pi)^3} \,\frac{1}{4M^2_B} p'_{D^*} \widetilde{p}_\nu \sum |t|^2  \, ,
\end{eqnarray}
where $p'_{D^*}$ is the $D^*$ momentum in the $B$ rest frame and $\widetilde{p}_\nu$ the  $\bar\nu$ momentum in the $\nu l$ rest frame,
  \renewcommand{\theequation}{4}
\begin{eqnarray}
p'_{D^*}=\frac{\lambda^{1/2}(m^2_B,M_{\rm inv}^{2(\nu l)}, m^2_{D^*})}{2 m_B}  \,, \quad
\widetilde{p}_\nu=\frac{\lambda^{1/2}(M_{\rm inv}^{2(\nu l)}, m^2_\nu,m^2_l)}{2 M_{\rm inv}^{(\nu l)}} \,.
\end{eqnarray}

  After considering  right-handed quark currents in terms of $\alpha$, we find for the different helicity contributions $(M'=0,\pm 1)$
\begin{itemize}
\item[1)] $M'=0$
 \renewcommand{\theequation}{5-a}
\begin{eqnarray}\label{neq:tM0}
 \sum |t|^2  &=&\frac{m^2_l}{m_{\nu} m_l}  \frac{M_{\rm inv}^{2(\nu l)}-m^2_l}{M_{\rm inv}^{2(\nu l)}} \big\{A A' (B+B')p \big\}^2 \blue{\alpha^2} \\ \nonumber \,
&+&  \frac{2}{m_{\nu} m_l}  \,\left(\widetilde{E}_\nu \widetilde{E}_l +\frac{1}{3} \widetilde{p}_\nu^2 \right) \big\{A A' (1+B B' p^2) \big\}^2 \blue{\alpha^2}   \, .
 \end{eqnarray}
 \item[2)] $M'=1$
  \renewcommand{\theequation}{5-b}
\begin{eqnarray}\label{neq:tM1}
\sum |t|^2 &= &\frac{2}{m_{\nu} m_l}  \,\left(\widetilde{E}_\nu \widetilde{E}_l +\frac{1}{3} \widetilde{p}_\nu^2 \right)
 \big\{A A' [(1-B B'p^2)\blue{\alpha} +(B p-B' p)]\big\}^2    \, .
 \end{eqnarray}
 \item[3)] $M'=-1$
 \renewcommand{\theequation}{5-c}
\begin{eqnarray}\label{neq:tMm1}
 \sum |t|^2 &=& \frac{2}{m_{\nu} m_l}  \,\left(\widetilde{E}_\nu \widetilde{E}_l +\frac{1}{3} \widetilde{p}_\nu^2 \right)
  \big\{A A' [(1-B B'p^2)\blue{\alpha} -(B p-B' p)]\big\}^2    \, .
 \end{eqnarray}\end{itemize}
Following the approach of \cite{epjc78},  the above the matrix elements are evaluated in the frame where  the   ${\bar\nu}  l$  system  is at rest, where  ${\bm{p}}_B={\bm{p}}_{D^*}={\bm{p}}$,  with $p$ given by
  \renewcommand{\theequation}{6}
\begin{eqnarray}
p = \frac{\lambda ^{1/2} (m_{B}^2, M_{\rm inv}^{2(\nu l)},  m^2_{D^*})}{2 M_{\rm inv}^{(\nu l)}} \, ,
\label{eq:p}
\end{eqnarray}
where $M_{\rm inv}^{(\nu l)}$ is the invariant mass of the $\nu l$ pair. For $B$ and $D^*$ mesons we have
  \renewcommand{\theequation}{7}
\begin{eqnarray}\label{eq:wfn2}
 A= \sqrt{\frac{E_B+m_B}{2 m_B}} \,,\quad  B= \frac{1}{m_B+E_B} \,; \quad  A'= \sqrt{\frac{E_{D^*}+m_{D^*}}{2 m_{D^*}}} \,,\quad  B'= \frac{1}{m_{D^*}+E_{D^*}} \,;
\end{eqnarray}
and $\widetilde{E}_\nu$, $\widetilde{E}_l$ are the  energies  of  $\bar{\nu}$ and  lepton, respectively,
  \renewcommand{\theequation}{8}
\begin{eqnarray}
\widetilde{E}_\nu=\frac{M_{\rm inv}^{2(\nu l)}+ m^2_\nu -m^2_l}{2\,M_{\rm inv}^{(\nu l)}}  \,,\qquad
\widetilde{E}_l=\frac{M_{\rm inv}^{2(\nu l)}+ m^2_l-m^2_\nu}{2\,M_{\rm inv}^{(\nu l)}}  \,.
\end{eqnarray}

\section{Results}

We present results of  $d \Gamma/d M_{\rm inv}^{(\nu l)}(\nu l\equiv  \bar{\nu}_\tau \tau)$ for different $M'=\pm 1,0$  and the sum
  \renewcommand{\theequation}{9}
\begin{eqnarray}\label{eq:R}
 R=\frac{d \Gamma}{d M_{\rm inv}^{(\nu l)}}|_{M'=0} + \frac{d \Gamma}{d M_{\rm inv}^{(\nu l)}}|_{M'=-1} +\frac{d \Gamma}{d M_{\rm inv}^{(\nu l)}}|_{M'=+1} \, . ~~~~~
\end{eqnarray}

\begin{figure}[ht]
\includegraphics[scale=0.8]{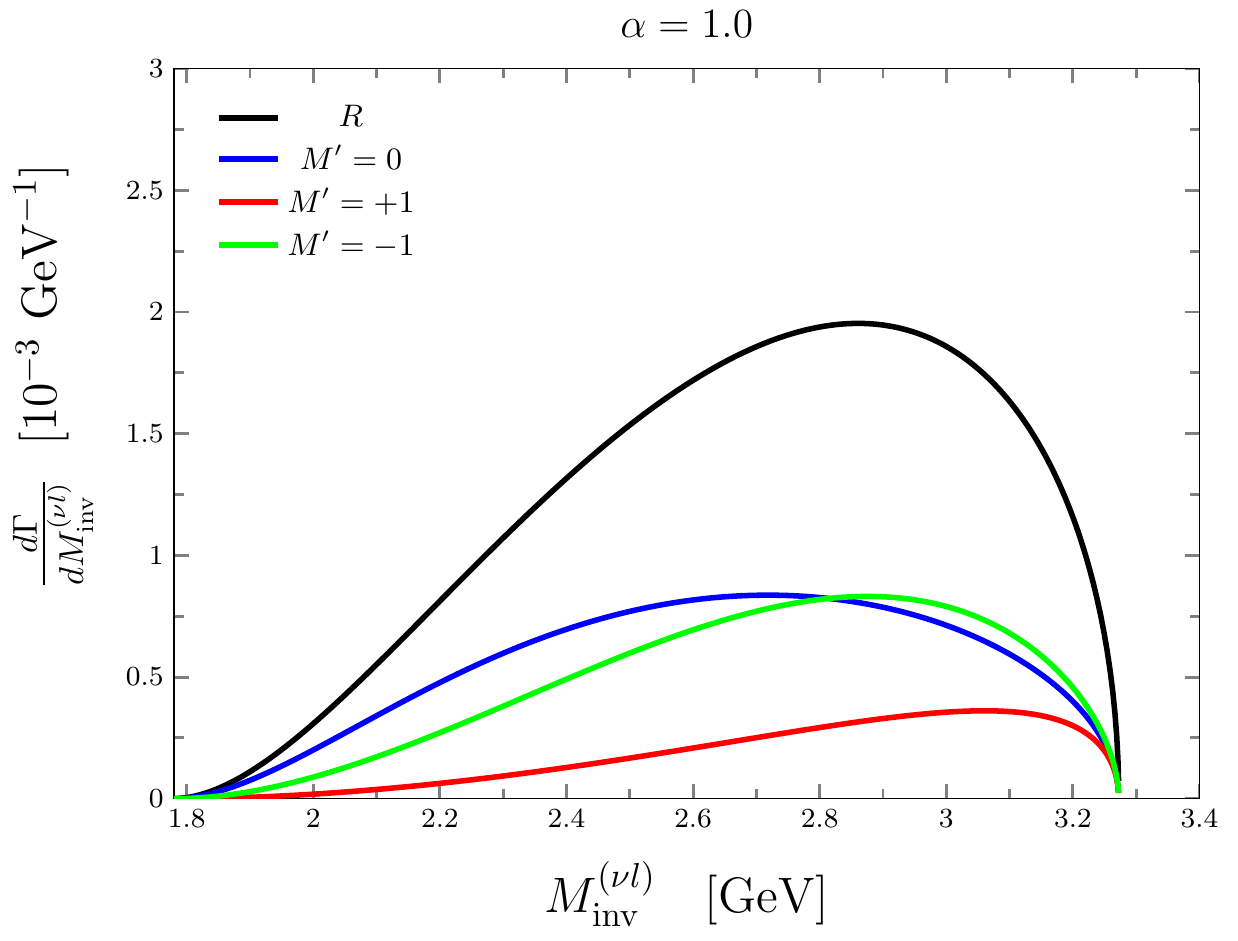}
\caption{ Total  differential width and individual contributions for different $M'=\pm 1,0$ components.}
% $\frac{d \Gamma}{d M_{\rm inv}^{(\nu l)}}|_{M'=0}$, $\frac{d \Gamma}{d M_{\rm inv}^{(\nu l)}}|_{M'=-1}$, and $\frac{d \Gamma}{d M_{\rm inv}^{(\nu l)}}|_{M'=+1}$.}
\label{fig:dg}
\end{figure}

\begin{figure}[ht!]
\includegraphics[scale=0.8]{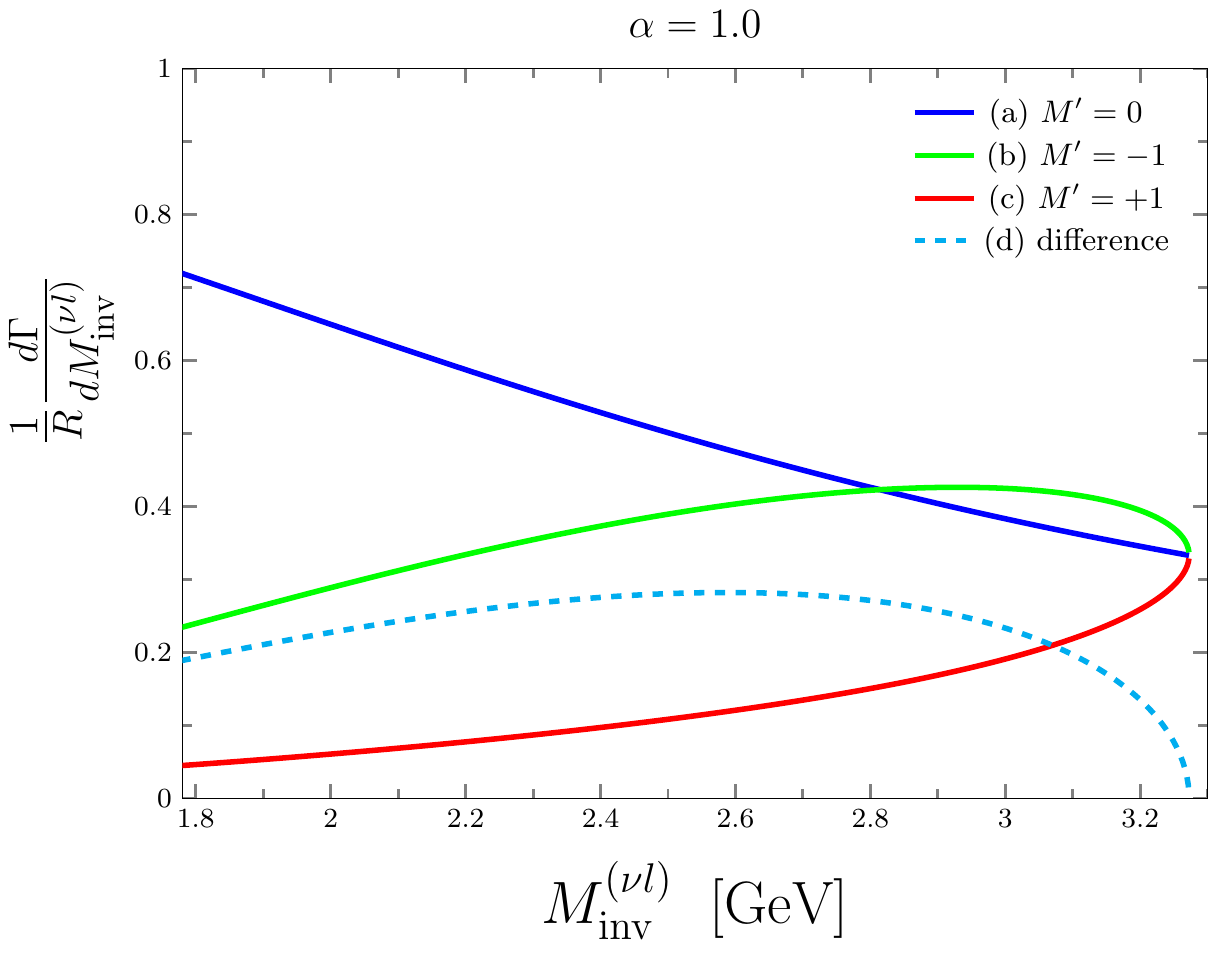}
\caption{The same as Fig. \ref{fig:dg} but normalized to the total.}
% The  lines (a), (b) and (c) show  $\frac{d \Gamma}{d M_{\rm inv}^{(\nu l)}}|_{M'=0}$, $\frac{d \Gamma}{d M_{\rm inv}^{(\nu l)}}|_{M'=-1}$, and
%$\frac{d \Gamma}{d M_{\rm inv}^{(\nu l)}}|_{M'=+1}$ respectively, and line (d) denotes the  difference of  $\frac{d \Gamma}{d M_{\rm inv}^{(\nu l)}}|_{M'=-1}-\frac{d \Gamma}{d M_{\rm inv}^{(\nu l)}}|_{M'=+1}$, all  divided by %the total differential width $R$ of Eq. \eqref{eq:R}. }
\label{fig:ro}
\end{figure}

In Fig. \ref{fig:dg} we  show the results for the different  contributions of $\alpha=1$ (SM) and  in Fig. \ref{fig:ro} we show the same results but normalized to the total.

\begin{figure}[ht!]
\includegraphics[scale=0.8]{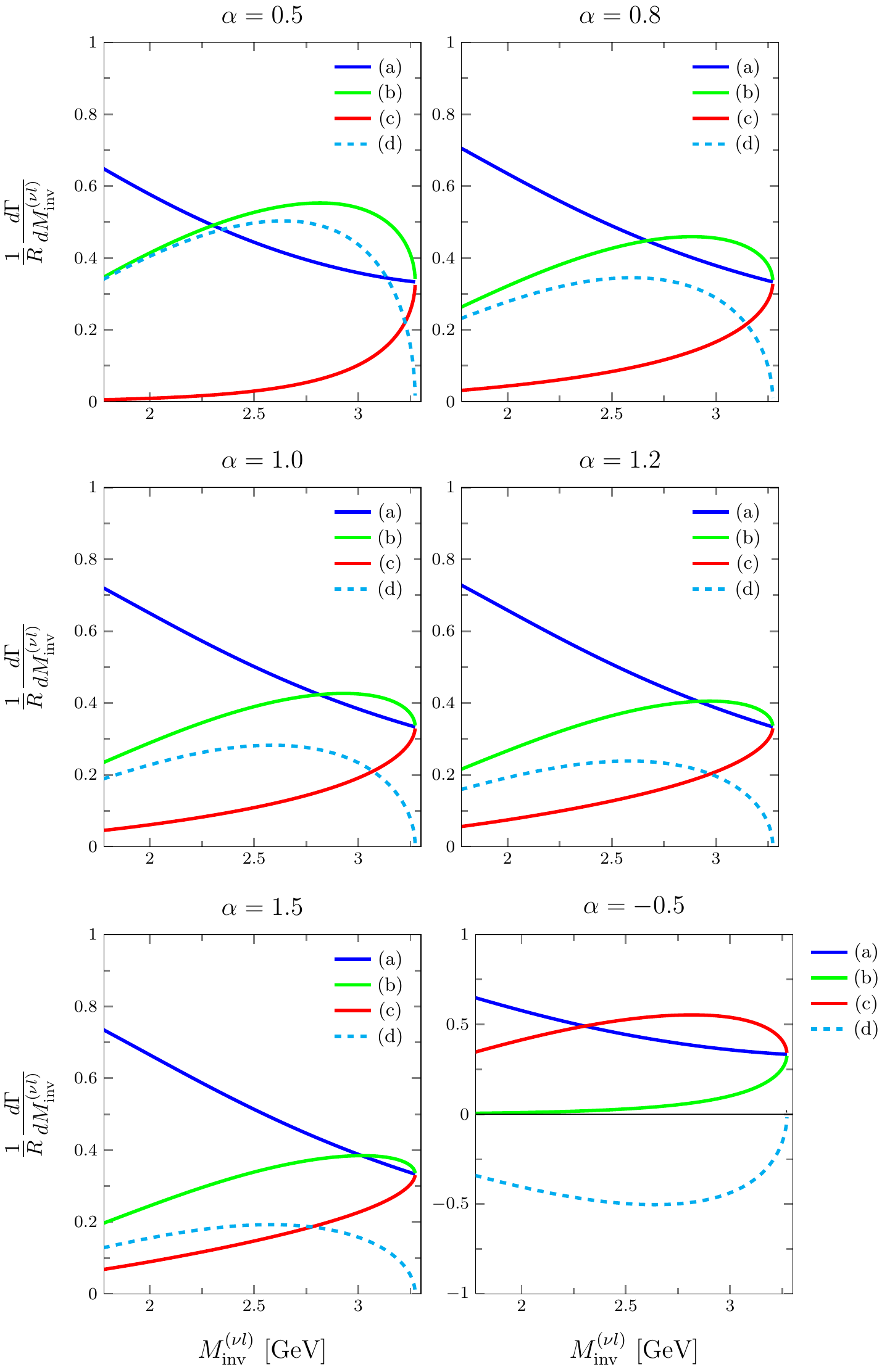}
\caption{The same as Fig. \ref{fig:ro} but for different $\alpha$.  }
\label{fig:aro}
\end{figure}

In Fig. \ref{fig:aro} we show the results of Fig. \ref{fig:ro} for different values of $\alpha$.  We can see that  all magnitudes change with $\alpha$, but the most spectacular is the difference
between the $M'=-1$ and $M'=+1$ contributions, which  change sign when $\alpha$ changes sign.

The magnitude $F_L^{D^*}$  is obtained integrating  $d \Gamma/d M_{\rm inv}^{(\nu l)}$  over the $M_{\rm inv}^{(\nu l)}$ variable and dividing by the total $\Gamma$.  We show the results for $F_L^{D^*}$  in Table \ref{tab:fd} for different values of $\alpha$.
\begin{table}[h!]
\renewcommand\arraystretch{1.}
\caption{The fraction of $D^{*}$ longitudinal polarization $f_L^{D^*}$ with different values of $\alpha$}
\centering
%\scriptsize\footnotesize\large
\begin{tabular}{c|ccccccc }
\toprule[1.0pt]
 $\alpha$ &~~~~~ $0.5$~~~~~  & ~~~~~ $0.8$~~~~~ & ~~~~~ $1.0$~~~~~ & ~~~~~ $1.2$~~~~~ & ~~~~~ $1.5$~~~~~ & ~~~~~ $-0.5$~~~~~    \\
\hline
$f_L^{D^*}$ & $0.415$  &  $0.448$ &  $0.456$ & $0.461$  &  $0.465$ &  $0.415$\\
\hline
\midrule[1.0pt]
\end{tabular}
\label{tab:fd}
\end{table}
We can see that for $\alpha=1$ (SM) we obtain
  \renewcommand{\theequation}{9}
\begin{eqnarray}
F_L^{D^*}=0.456 \,,
\end{eqnarray}
which is in  remarkable agreement with the result of \cite{srimoy}. We  should note  that in \cite{epjc78} no  free parameters nor fit to data are used, but  ratios are expected to be relatively accurate.
The results in Table \ref{tab:fd}  are interesting because we observe that $F_L^{D^*}$ is very insensitive to the value of $\alpha$.  We can claim that for this family of models, $F_L^{D^*}$ is not a good
magnitude to test contributions beyond SM.

\begin{figure}[ht!]
\includegraphics[scale=0.8]{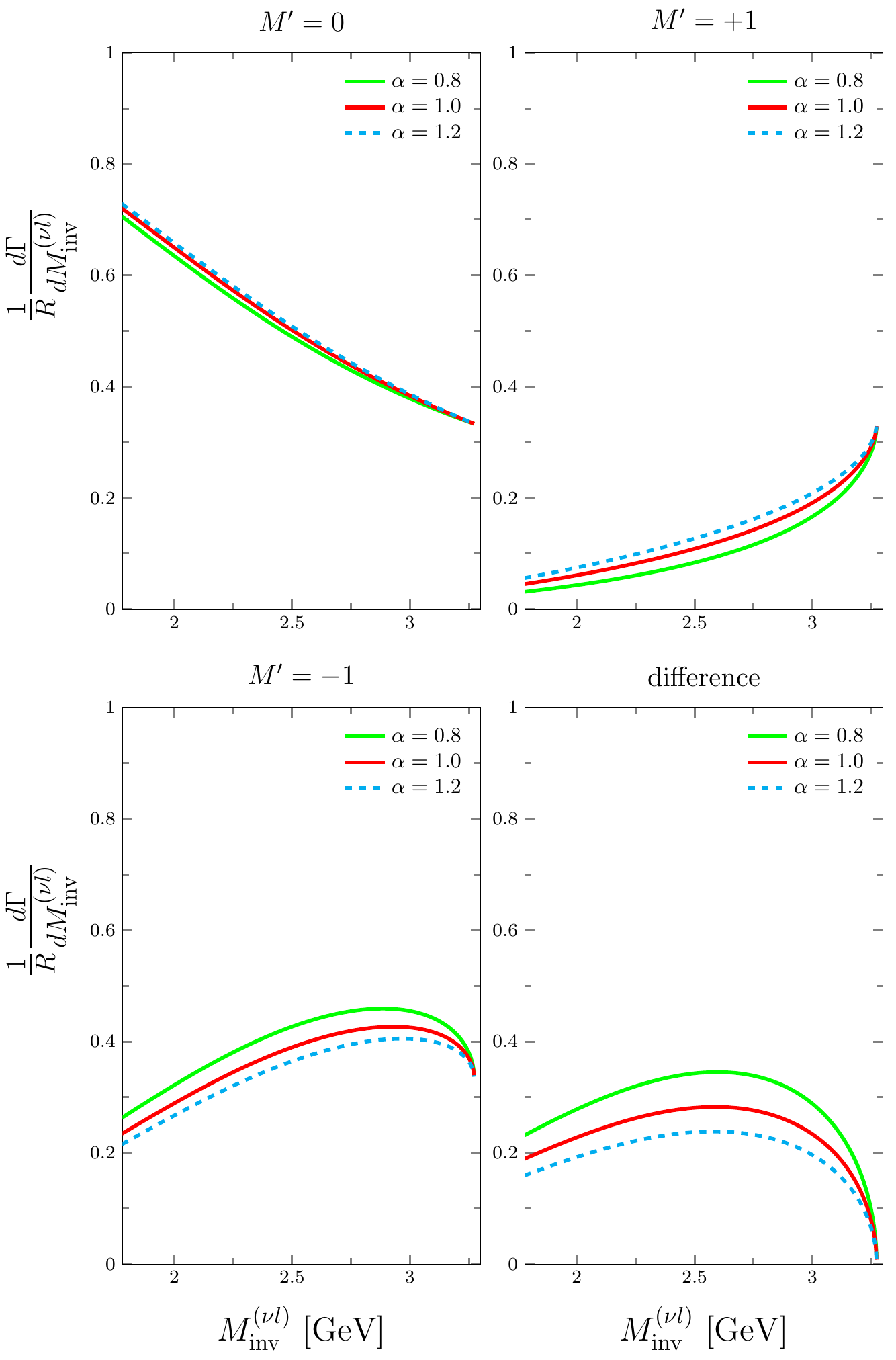}
\caption{
%The same as Fig. \ref{fig:ro} but for different $\alpha$.
  The longitudinal $M'=0$,  two transverse $M'=\pm 1$ components  and the difference for the range of $\alpha \in [0.8-1.2]$. }
\label{fig:aM}
\end{figure}

On the other hand  we see that the measurement of the two transverse components carries more information and we provide this information in Fig. \ref{fig:aM}, in which
we show the results obtained for the different ratios, and the difference of $M'=-1$  and $M'=+1$ for different values of $\alpha$.  One can see that
the longitudinal component $M'=0$ is less sensitive than any of the other two, and in particular the difference between the $M'=-1$  and $M'=+1$ components
is the most sensitive magnitude.  If we take the range of value $\alpha \in [0.8-1.2]$ the band of values for the  $M'=0$ contribution is quite narrow, while the band for
the difference between the $M'=-1$  and $M'=+1$ components is considerably larger.

\section{Summary}

As a summary, we have shown that the model used, which evaluates explicitly the operators and contains  no free parameters, is in remarkable agreement for the value of $F_L^{D^*}$
with the most sophisticated evaluations of the SM.  In view of this, we extended  the   model to calculate the contributions of the longitudinal  and transverse helicities  of the
$\bar{B} \to  D^{*} \bar{\nu}_\tau \tau$  reaction for a family of models  BSM  with right handed quark currents.
  We concluded  that the measurement of the transverse helicity components in the $\bar{B} \to  D^{*} \bar{\nu}_\tau \tau$ reaction is a more promising tool than the
longitudinal helicity in the search for potential extrapolations  of the SM and strongly suggest to study these components  experimentally.

\section*{Acknowledgments}
We thank L. S. Geng for useful discussions on the work. LRD acknowledges the support from the National Natural Science Foundation of China (Grant Nos. 11975009, 11575076).  This work is partly supported by the Spanish Ministerio
de Economia y Competitividad and European FEDER funds under Contracts No. FIS2017-84038-C2-1-P B
and No. FIS2017-84038-C2-2-P B, and the Generalitat Valenciana in the program Prometeo II-2014/068, and
the project Severo Ochoa of IFIC, SEV-2014-0398 (EO).

\end{document}